\documentclass[%
 reprint,
 superscriptaddress,
 amsmath,amssymb,
 aps,
 pra,
]{revtex4-2}

\usepackage{graphicx} 
\usepackage{bm}
\usepackage[colorlinks,urlcolor=blue,citecolor=blue,linkcolor=blue,pdfstartview=FitH]{hyperref}
\usepackage{enumerate}
\usepackage{amsmath}
\usepackage{color}
\usepackage{cases}
\usepackage{microtype}
\usepackage{siunitx}
\usepackage{todonotes}
\usepackage[utf8]{inputenc}
\usepackage[T1]{fontenc}
\usepackage{txfonts}           

\newcommand{\diff}{\mathrm{d}}
\newcommand{\kcut}{k_{\mathrm{cut}}}
\newcommand{\Lc}{L_\mathrm{c}}
\newcommand{\nv}{n_\mathrm{v}}

\begin{document}

\title{Universal coarsening of a two-dimensional Bose gas under conservative evolution}

\author{A. J. Groszek}
\affiliation{Australian Research Council Centre of Excellence for Engineered Quantum Systems, School of Mathematics and Physics, University of Queensland, St. Lucia, QLD 4072, Australia.}

\author{T. P. Billam}
\affiliation{Joint Quantum Centre (JQC) Durham--Newcastle, School of Mathematics, Statistics and Physics, Newcastle University, Newcastle upon Tyne, NE1 7RU, United Kingdom}

\begin{abstract}
We investigate the phase ordering dynamics of a uniform two-dimensional Bose gas quenched to a finite temperature in the superfluid phase. Starting from a defect-rich, far-from-equilibrium state, we model the subsequent evolution with the projected Gross--Pitaevskii equation, which conserves both energy and particle number. By tuning the initial energy, we control the effective post-quench temperature and examine its role in the equilibration dynamics. We find that the gas exhibits universal behaviour at all temperatures, evidenced by spatio-temporal scaling of correlation functions and power-law growth of the correlation length $\sim t^{1/z}$, with $z$ the dynamical critical exponent. We find $z$ to be temperature dependent, with $z \approx 1.5$ for post-quench temperatures just below the Berezinskii--Kosterlitz--Thouless (BKT) transition, and $z \approx 1.9$ for quenches to near-zero temperature. Analysis of the Porod tail of the momentum distribution suggests a temperature-dependent competition between vortices and sound waves in the coarsening process. The two-time correlation function also exhibits universal scaling, decaying as $\sim t^{-\lambda/z}$, with autocorrelation exponent $\lambda$. Near the BKT transition we obtain $\lambda \approx 2$, whereas $\lambda$ is found to diverge as the effective temperature approaches zero.
\end{abstract}

\maketitle

\section{Introduction}

Many-body systems driven far from equilibrium can exhibit universal dynamical scaling as they relax towards a steady state. In the scaling regime, statistical properties of the system become unchanged in time, except for an overall change in lengthscale. This self-similar behaviour is revealed by spatio-temporal scaling of correlations, with associated power-law exponents that are expected to identify the system's dynamical universality class~\cite{universality_footnote}. These exponents should depend only on global properties such as dimensionality and underlying symmetries, and not on microscopic details of the quench. Universal scaling of this type has been firmly established in classical models including the Ising model~\cite{humayun_non-equilibrium_1991, kim_nonequilibrium_2003}, nematic fluids~\cite{zapotocky_kinetics_1995}, and the XY model~\cite{yurke_coarsening_1993, rojas_dynamical_1999, bray_breakdown_2000, jelic_quench_2011}.

The theory of phase ordering kinetics~\cite{bray_theory_1994} describes universal scaling in terms of domain coarsening, whereby localised patches of the new equilibrium phase form immediately after the quench and grow in time in a self-similar way, gradually merging via the annihilation of topological defects. Central to this theory is the prediction that the correlation length $\Lc$ grows in time as a power-law, $\Lc \sim t^{1/z}$, with $z$ the dynamical critical exponent. This behaviour can also be understood from the perspective of renormalisation group theory, where the power-law growth of correlations reflects a critical slowing down that occurs when the system passes by a fixed point in the `system space' during its evolution~\cite{calabrese_ageing_2005}. Recently, there has been particular interest in understanding universal dynamics within this framework~\cite{schmied_non-thermal_2019, mikheev_universal_2023}, as it highlights the possibility that multiple fixed points may influence the dynamics, each with their own scaling properties and associated critical exponents.

Ultracold atomic Bose gases provide an ideal setting for exploring universal coarsening dynamics, owing to their high level of controllability. Indeed, experiments to date have demonstrated universal dynamics in scalar Bose gases across one~\cite{erne_universal_2018}, two~\cite{sunami_universal_2023, gazo_universal_2025, chang_coupling-induced_2025} and three dimensions~\cite{glidden_bidirectional_2021, garcia-orozco_universal_2022, martirosyan_universal_2025}, as well as in spinor Bose gases in one~\cite{prufer_observation_2018, lannig_observation_2023} and two dimensions~\cite{huh_universality_2024}. Closely related studies have also explored vortex and wave turbulence~\cite{navon_emergence_2016, gauthier_giant_2019, johnstone_evolution_2019, navon_synthetic_2019, galka_emergence_2022, dogra_universal_2023} and the Kibble--Zurek mechanism~\cite{weiler_spontaneous_2008, lamporesi_spontaneous_2013, chomaz_emergence_2015, navon_critical_2015, goo_defect_2021, goo_universal_2022}, making universal scaling an important theme of current experimental research. Coarsening has also been extensively explored via numerical simulations, with particular interest in two-dimensional (2D) scalar~\cite{damle_phase_1996, karl_strongly_2017, comaron_quench_2019, groszek_crossover_2021}, binary~\cite{hofmann_coarsening_2014, wheeler_relaxation_2021} and spinor~\cite{williamson_universal_2016, williamson_coarsening_2017, symes_nematic_2017, williamson_anomalous_2019} Bose gases, as well as driven--dissipative systems~\cite{kulczykowski_phase_2017, comaron_dynamical_2018}.

In Ref.~\cite{groszek_crossover_2021}, we investigated the universal coarsening of a two-dimensional (2D) Bose gas following an instantaneous quench into the superfluid phase. We fixed the post-quench temperature and varied the dissipation strength, finding that in general the dynamical critical exponent significantly deviates from the typically expected value of $z=2$ for this system~\cite{chantesana_kinetic_2019, gazo_universal_2025}. In the limit of strong dissipation (known as `Model A'~\cite{hohenberg_theory_1977}) we found $z>2$, a value generally considered to be consistent with $z=2$, provided that logarithmic corrections to scaling are accounted for~\cite{yurke_coarsening_1993, rutenberg_energy-scaling_1995}. By contrast, in the limit of zero dissipation (i.e.~conservative evolution) we found that $z<2$, which cannot be explained by the same corrections. We suggested that this behaviour arose from a vortex `mobility' that itself grows as a power-law in time during the scaling regime~\cite{groszek_crossover_2021}. Two recent 2D Bose gas experiments have also found $z \approx 2$~\cite{gazo_universal_2025} and $z = 1.73(9)$~\cite{chang_coupling-induced_2025}, respectively, in broad agreement with our findings in this regime.

Here, we expand on the results of Ref.~\cite{groszek_crossover_2021}, focusing on the conservative limit. We analyse the scaling behaviour of several additional observables, and explore the effect of the final post-quench temperature on the measured exponents. The main results of this work are presented in Fig.~\ref{fig:z_vs_eta}. In summary, we obtain strong evidence of universal coarsening regardless of temperature, but find that the exponent $z$ exhibits temperature dependence. In particular, $z \approx 1.5$ for quenches just below the Berezinskii--Kosterlitz--Thouless (BKT) critical point~\cite{berezinskii_destruction_1971, berezinskii_destruction_1972, kosterlitz_ordering_1973}, while $z \approx 1.9$ for quenches to near-zero temperatures. We also observe power-law decay of the autocorrelation function $\sim t^{-\lambda/z}$, with the autocorrelation exponent $\lambda$ likewise exhibiting strong temperature dependence. Surprisingly, $\lambda$ appears to diverge as the temperature approaches zero.

The paper is structured as follows. In Sec.~\ref{sec:setup} we describe the system setup, including a description of the model, parameter choices, and our procedure for generating initial conditions. Section~\ref{sec:quench_example} contains a detailed analysis of the universal scaling dynamics of the system at a single temperature. We characterise the dynamics using five key measures: (i) the spatial correlation function, (ii) the vortex density (iii) the momentum distribution, (iv) finite-size scaling of the condensate fraction, and (v) the autocorrelation function. In Sec.~\ref{sec:Tdependence} we then present our main results, which we have obtained by applying these methods of analysis across a range of temperatures. We conclude in Sec.~\ref{sec:conclusion}.

\section{System setup and numerics\label{sec:setup}}

We simulate the dynamics of a two-dimensional Bose gas at nonzero temperature using the classical field methodology. Within this framework, the gas is represented as a complex scalar field $\psi(\textbf{r}, t)$, which includes contributions from all highly occupied single-particle energy states of the system, up to some chosen cutoff in the single-particle energy spectrum. We model the evolution of this field using the projected Gross--Pitaevskii equation (PGPE)~\cite{davis_simulations_2001, blakie_dynamics_2008}:
\begin{equation} \label{eq:PGPE}
    i \hbar \frac{\partial \psi}{\partial t} = \mathcal{P} \left \lbrace - \frac{\hbar^2}{2m} \nabla^2 \psi + g \left | \psi \right | ^2 \psi \right \rbrace.
\end{equation}
In this expression, the projection operator $\mathcal{P}$ enforces the energy cutoff by preventing particle transfer outside the chosen subset of energy modes. The parameters $m$ and $g$ correspond to the particle mass and the two-dimensional interaction strength, respectively. Equation~\eqref{eq:PGPE} describes a closed system, for which both energy
$E = \int ( \hbar^2 |\nabla \psi|^2 / 2m + g | \psi |^4 / 2 ) \diff \mathbf{r}$
and particle number
$N = \int | \psi |^2 \diff \mathbf{r}$ are conserved under time evolution.

We consider a system in a doubly periodic square domain of size $L \times L$, and represent the field using a plane wave basis satisfying $|\textbf{k}| < \kcut$, where $\kcut$ is our chosen wavenumber cutoff. We initiate the wavefunction by populating a disk of radius $k_{\rm d}$ in wavenumber space, $\psi = \sum_{|\textbf{k}| < k_{\rm d}} \sqrt{n_\textbf{k}} \exp{[i (\textbf{k} \cdot \textbf{r} + \phi_\textbf{k})]}$, which ensures zero linear momentum. The late-time equilibrium properties of the system are therefore fully determined by the energy $E$ and particle number $N$; these two quantities determine the temperature of the system in this microcanonical ensemble~\cite{davis_simulations_2001, blakie_dynamics_2008}. The populations $n_\textbf{k}$ are set to be uniform across all modes, enforcing a chosen mean density $n \equiv N / L^2$. Given this constraint, the phases $\phi_\textbf{k}$ then determine the (mean) energy density $\epsilon \equiv E / L^2$. To obtain a target $\epsilon$, we first randomly sample each $\phi_\textbf{k}$ uniformly in the interval $[0,2\pi)$. We then use a Powell minimisation algorithm~\cite{powell_efficient_1964} to adjust each phase until the desired $\epsilon$ is reached, within a relative precision of $\mathcal{O} ( 10^{-5} )$. This process produces far-from-equilibrium initial states at a desired $\epsilon$ and $n$, containing a high density of quantised vortices and antivortices.

Throughout this work, we express dimensionful quantities in terms of the chemical potential $\mu = g n$, the healing length $\xi = \hbar / (m \mu)^{1/2}$ and the timescale $\hbar / \mu$. We choose interaction strength $g = 0.3 \, \hbar^2 / m$ and wavenumber cutoff $k_{\rm cut} = \pi / (2 \Delta x)$, where the numerical grid spacing $\Delta x \approx \xi / 2$. Our choice of cutoff ensures that the field remains de-aliased~\cite{blakie_dynamics_2008}. We fix the mean particle density to $n \approx 3.1 \, \xi^{-2}$, and vary the energy density between $\epsilon \approx 1.8 \, \mu \xi^{-2}$ and $\epsilon \approx 3.6 \, \mu \xi^{-2}$, allowing us to perform quenches across a range of equilibrium (microcanonical) temperatures across the superfluid regime. At each $\epsilon$, the radius $k_\mathrm{d}$ of initially populated modes is made as large as possible while still ensuring that the minimiser can obtain states. To this end, $k_{\rm d}$ is increased monotonically with $\epsilon$ in the range $0.28\, \xi^{-1} \lesssim k_\mathrm{d} \lesssim 1.23 \, \xi^{-1}$. Unless otherwise stated, we use a system size of $L \approx 262 \, \xi$ and ensemble average the results over $\mathcal{N}=256$ realisations of statistically equivalent initial states.

Equation~\eqref{eq:PGPE} is solved numerically using a fourth-order adaptive Runge--Kutta scheme, optimised for performance on a graphics processing unit using {\small CUDA}~\cite{cuda_toolkit}. The energy and particle number are conserved to precisions of $\mathcal{O}(10^{-5})$ and $\mathcal{O}(10^{-6})$, respectively. In Sec.~\ref{sec:Tdependence}, while the PGPE is still used for dynamics, we also make use of the stochastic projected Gross--Pitaevskii equation~\cite{blakie_dynamics_2008} in order to reduce the computational cost of sampling equilibrium states (for details, see Appendix~\ref{app:spgpe_equilibrium}).

\begin{figure}[t]
    \centering
    \includegraphics[width=\columnwidth]{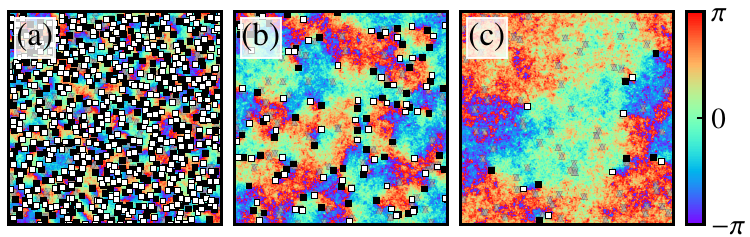}
    \caption{Evolution of the phase of the field $\psi$ following a quench, with mean energy density $\epsilon \approx 2.8 \, \mu \xi^{-2}$. From left to right, the frames correspond to times $\mu t / \hbar = \lbrace 100, 1000, 5000 \rbrace$, respectively. White (black) squares denote vortices (antivortices) that have been detected after passing the field through a low-pass filter with $k_{\rm f} \approx 0.92 \, \xi^{-1}$, while grey upright (inverted) triangles denote thermally activated vortices (antivortices) detected in the raw field (see Sec.~\ref{sub:vortex_decay}).}
    \label{fig:phase_vortices}
\end{figure}

\section{Universal dynamics following a quench \label{sec:quench_example}}

In this section, we present results from a quench with a mean energy density $\epsilon \approx 2.8 \, \mu \xi^{-2}$, corresponding to an equilibrium temperature deep within the superfluid regime. The initial states are obtained by populating wavenumbers up to $k_\mathrm{d} \approx 0.92 \, \xi^{-1}$ (as described in the previous section). Figure~\ref{fig:phase_vortices} illustrates the evolution of the system following the quench. The phase of the field, $\arg \lbrace \psi \rbrace$, is shown at three times, with vortices and antivortices identified as white and black squares, respectively.

We find strong evidence of universal scaling behaviour during the dynamics as the defects annihilate and the system approaches equilibrium. In the following, we analyse the dynamical scaling behaviour of five independently measured observables: (i) correlation functions (Sec.~\ref{sub:correlation_length}), (ii) vortex density (Sec.~\ref{sub:vortex_decay}), (iii) momentum distributions (Sec.~\ref{sub:momentum_spectrum}), (iv) the condensate population (Sec.~\ref{sub:condensate_Geq}), and (v) autocorrelation functions (Sec.~\ref{sub:autocorrelation}).

\subsection{Growth of the correlation length \label{sub:correlation_length}}

To measure the spatial coherence of the field, we follow the approach described in Ref.~\cite{groszek_crossover_2021}. Specifically, we calculate the first-order (equal-time) correlation function,
\begin{equation} \label{eq:correlation_function}
G(\mathbf{r},t) = \frac{\langle \psi^*(\mathbf{r}+\mathbf{r}', t) \psi(\mathbf{r}', t)  \rangle}{\sqrt{ \langle | \psi(\mathbf{r}+\mathbf{r}', t) |^2 \rangle \langle | \psi(\mathbf{r}', t) | ^2 \rangle}},
\end{equation}
where the angular brackets denote an average over both the statistical ensemble and the coordinate $\mathbf{r}'$. Numerically, this quantity can be calculated efficiently via the Wiener--Khinchin theorem.

The scaling hypothesis predicts that a system undergoing universal dynamics should exhibit correlations of the form
\begin{equation} \label{eq:scaling_hypothesis}
    G(r, t) = G_{\rm eq}(r) F(r, t),
\end{equation}
where $G_{\rm eq}(r) = G(r, t \to \infty)$ is the equilibrium correlation function, and $F$ is a scaling function that obeys $F(r,t) = F(r/\Lc(t))$, with $F(0)=1$~\cite{bray_theory_1994}. The correlation length $\Lc(t)$ corresponds to the mean lengthscale over which the system has locally reached equilibrium, and can be thought of as the average size of the phase domains at time $t$. In words, the scaling hypothesis posits that the correlations in the scaling regime attain a time-independent form, up to a rescaling of lengthscales by $\Lc(t)$. Simultaneously, this lengthscale should itself grow as a power-law, $\Lc(t) \sim t^{1/z}$.

For a 2D Bose gas below the BKT critical point, the equilibrium correlation function in Eq.~\eqref{eq:scaling_hypothesis} decays algebraically as $G_{\rm eq}(r) \sim r^{-\eta}$ for $r \gg \xi$ in the thermodynamic limit, with temperature-dependent exponent $0 \leq \eta \leq 0.25$~\cite{pethick_bose-einstein_2008}. In the numerics, we obtain $G_{\rm eq}(r)$ by averaging over $\mathcal{O}(10^3)$ states sampled from the PGPE at late times after the system has reached equilibrium. For the energy density $\epsilon \approx 2.8\, \mu \xi^{-2}$ we are considering in this section, we obtain an exponent $\eta \approx 0.106$, confirming that this configuration is deep within the superfluid phase (see Sec.~\ref{sub:condensate_Geq} for details). With the equilibrium properties known, we can use Eq.~\eqref{eq:scaling_hypothesis} to calculate the scaling function $F(r,t)$ throughout the preceding coarsening dynamics, defining the correlation length $\Lc(t)$ to be the lengthscale satisfying $F(\Lc,t) = F_0$ for some chosen threshold value $F_0$. Here we use $F_0=0.5$, as this choice should minimise both discretisation and finite-size effects (at small and large scales, respectively). 

\begin{figure}[t]
    \centering
    \includegraphics[width=\columnwidth]{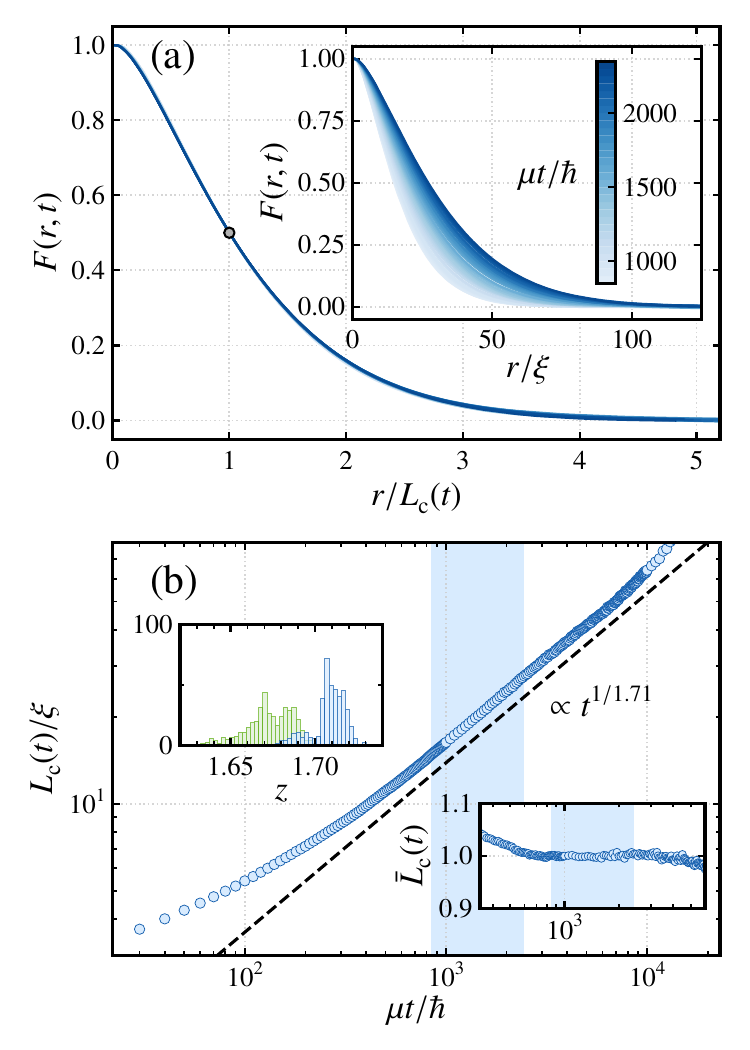}
    \caption{(a) Ensemble-averaged scaling function $F(r,t)$ before (inset) and after (main frame) rescaling the radial distance $r$ by the correlation length $\Lc(t)$. The colour bar indicates the time at which each curve has been sampled, and the grey dot denotes the threshold $F_0=0.5$ used to define $\Lc$. (b) Evolution of the correlation length $\Lc(t)$, with the scaling window highlighted, and the best power-law fit to the data shown as a black dashed line (offset for visibility). The right inset of (b) shows the compensated correlation length $\bar{L}_\mathrm{c}(t)$, obtained by dividing $\Lc(t)$ by the power-law fit (horizontal axis same as main frame). The left inset shows the distribution of exponents $z$ measured from fits to different subintervals within the scaling window. Blue/right peak corresponds to the averaged $F(r,t)$, while green/left peak corresponds to the ensemble-averaged $\Lc(t)$ (see text).}
    \label{fig:correlation_function}
\end{figure}

In a finite system, universal scaling dynamics will terminate once the correlation length approaches the system size, signalling the onset of equilibrium. In order to extract meaningful scaling exponents, we must therefore restrict our analysis to a temporal window that excludes both these late-time finite-size effects, as well as initial transients resulting from `memory' of the initial condition. To specify this scaling window, we first define its endpoint to be the earliest time for which $\Lc(t) > L/4$ in any one of the simulations in the ensemble. The starting point is then determined by extending the window back in time as far as possible while ensuring there are minimal deviations from the unique scaling function $F(r/\Lc(t))$ (we have previously quantified this method in Ref.~\cite{groszek_crossover_2021}). Using this approach, we identify the scaling window to be $850 \leq \mu t / \hbar \leq 2400$ for this energy density.

The evolution of the scaling function $F(r,t)$ over this temporal window is shown in Fig.~\ref{fig:correlation_function}(a). The raw data (inset) is seen to collapse onto a single curve upon rescaling $r \to r / \Lc(t)$ (main frame), providing strong evidence for universal scaling in this system. The correlation length is plotted in Fig.~\ref{fig:correlation_function}(b), and power-law scaling $\Lc \sim t^{1/z}$ is visible over the same temporal window (shaded region). To measure the exponent $z$, we fit a power-law to the data across all possible subintervals of $\geq 14$ consecutive points within the scaling window. This yields a distribution of $z$ values characterising the statistical uncertainty associated with the choice of fitting window [right peak in the left inset of (b)]. We measure $z$ and its uncertainty as the mean and standard deviation of this distribution, respectively, resulting in $z=1.71(1)$ for this quench. The corresponding power-law is shown as a black dashed line in (b). As an indicator of the goodness-of-fit, the compensated correlation length $\bar{L}_{\rm c}(t) = \Lc(t) / \Lc^{\rm fit}(t)$ is also plotted [right inset of (b)]. Within the scaling window, it remains close to one, indicating that the data is well described by the fit.

In the above, we have ensemble-averaged the scaling function $F(r,t)$ before measuring $\Lc(t)$, although we note that $\Lc(t)$ could alternatively be obtained from each individual simulation first, before performing ensemble averaging. Doing so yields an exponent $z=1.67(2)$, measured from the corresponding $z$-distribution shown in green in the left inset of Fig.~\ref{fig:correlation_function}(b). The close agreement of the two exponents indicates that the method of averaging has only a minimal effect on the final result. As noted in Ref.~\cite{groszek_crossover_2021}, the choice of threshold $F_0$ also has a weak systematic effect on the obtained value of $z$. For $F_0=0.3$, we obtain $z=1.68(1)$, and for $F_0=0.7$ we find $z=1.76(2)$.

\subsection{Decay of vortices \label{sub:vortex_decay}}

As the superfluid relaxes towards equilibrium, vortices created at the time of the quench gradually decay via vortex--antivortex annihilation events, allowing phase coherence to develop across the system (see Fig.~\ref{fig:phase_vortices}). For randomly and uniformly distributed vortices, we expect the correlation length to be on the order of the mean distance between vortices. This suggests a scaling law $\nv(t) \sim \Lc^{-2}(t) \sim t^{-2/z}$ for the vortex density $n_\mathrm{v}$, providing an independent approach for measuring the exponent $z$.

\begin{figure}
     \centering
     \includegraphics[width=1.0\columnwidth]{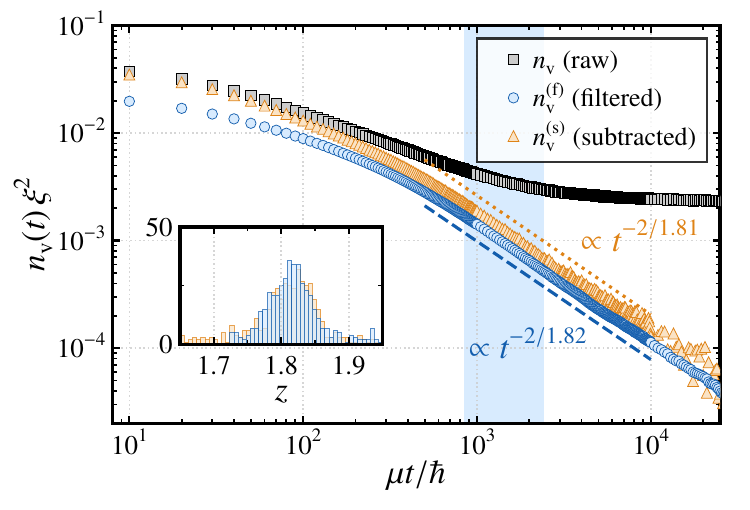}
     \caption{
     Evolution of the vortex density following the quench: raw vortex density $\nv$ (black squares), filtered free vortex density $\nv^\mathrm{(f)}$ (blue circles), and subtracted free vortex density $\nv^\mathrm{(s)}$ (orange triangles). The scaling window is highlighted, and power-law fits within this region are shown for $\nv^\mathrm{(f)}$ (dashed line) and $\nv^\mathrm{(s)}$ (dotted line). The lines are offset for clarity. The inset shows the distributions of exponents $z$ obtained from different choices of fitting window (colours match the legend). For reference, a vortex density of $\nv\xi^2 \approx 10^{-5}$ corresponds to a single vortex left in the system.}
     \label{fig:vortex_density}
 \end{figure}

We identify vortices and antivortices in the field $\psi(\textbf{r},t)$ by finding all points around which the phase winds by $\pm 2\pi$. The raw vortex density measured in this way is plotted in Fig.~\ref{fig:vortex_density} (black squares). Due to the strong thermal fluctuations present following the quench, a significant fraction of these vortices are tightly bound, thermally-activated vortex--antivortex pairs~\cite{simula_thermal_2006}. These thermal dipoles persist in the field even as the system equilibrates, as indicated by the late-time plateau in the data. The predicted scaling law is therefore obscured and cannot be seen in the raw vortex density. However, these thermal vortex pairs do not influence the phase profile at scales $\gtrsim \xi$, and hence have a minimal effect on $\Lc$. If they can be removed from consideration, the density of `free' vortices may in fact exhibit scaling.

Removing thermal dipoles is a nontrivial task, but we nonetheless try two approaches here. The first method is to apply a low-pass filter to the field $\psi$ in Fourier space by removing contributions from wavenumbers $|\textbf{k}| > k_{\rm f}$, for some chosen filtering wavenumber $k_{\rm f}$. We then perform the vortex detection and obtain the `filtered' free vortex density, which we denote $n_{\rm v}^{\rm (f)}(t)$. Figure~\ref{fig:phase_vortices} demonstrates the result of this process: black/white squares show the vortex configuration after filtering, while grey triangles identify vortices that have been removed by the filtering process (which always appear as closely bound dipoles). In the second approach, we measure the mean equilibrium vortex density $\tilde{n}_\mathrm{v}$ by averaging the raw vortex density over the latest times in our simulations, in this case obtaining $\tilde{n}_\mathrm{v} \approx 2.3 \times 10^{-3} \, \xi^{-2}$. Assuming that the thermal vortex density remains constant for all time, we then define the `subtracted' free vortex density $n_{\rm v}^{\rm (s)}(t) \equiv \nv(t) - \tilde{n}_\mathrm{v}$. We note that it is not clear that this assumption is valid when the system is far from equilibrium.

Figure~\ref{fig:vortex_density} shows both free vortex density estimates, $\nv^\mathrm{(f)}(t)$ (blue circles) and $\nv^\mathrm{(s)}(t)$ (orange triangles), which unlike the raw vortex density do exhibit power-law decay within the highlighted scaling window identified in the previous section. To measure $\nv^\mathrm{(f)}(t)$, we have chosen a filter of radius $k_{\rm f} \approx 0.92 \, \xi^{-1}$ in $k$-space. We find that this choice maximises the length of time for which $\nv^\mathrm{(f)}(t)$ follows a power-law. We fit $\nv(t) \sim t^{-2/z}$ to both datasets within the scaling window, using the same fitting method as for $\Lc(t)$. The resulting $z$-distributions are shown in the inset of Fig.~\ref{fig:vortex_density}, and from these we obtain $z^\mathrm{(f)}=1.82(4)$ and $z^\mathrm{(s)}=1.81(5)$. Despite the difference in measurement technique, these exponents agree well with one another, providing evidence that both approaches are valid. However, the obtained exponents are somewhat larger than that measured from the $\Lc(t)$ scaling. In Sec.~\ref{sec:Tdependence}, we repeat this analysis for other quench energies, and find that even at lower energies where thermal vortex pairs are absent, the $\nv$ exponents remain larger than the $\Lc$ exponent. This suggests that measuring $z$ from $\nv(t)$ gives rise to systematic errors, possibly due to a violation of the assumption that $\nv(t) \sim \Lc^{-2}(t)$. Despite this, $\nv(t)$ still predicts $z<2$ in general.

A number of works have described the decay of vortices in 2D Bose gases using phenomenological collision rate equations of the form $\diff \nv / \diff t \sim - [\nv(t)]^\zeta$~\cite{kwon_relaxation_2014, groszek_onsager_2016, cidrim_controlled_2016, karl_strongly_2017, baggaley_decay_2018, groszek_decaying_2020}. In this picture, it is argued that $\zeta$ is related to the number of vortices contributing to each annihilation event, and importantly, $\zeta = 2$ is predicted for two-vortex annihilation events. Assuming $\zeta > 1$, the solution of this differential equation is $\nv(t) \sim t^{-1 / (\zeta - 1)}$. Equating this to the earlier prediction $\nv(t) \sim t^{-2/z}$, the relation $\zeta = 1 + z/2$ can be inferred. Interestingly, taking $z<2$ as obtained here, we also must have $\zeta<2$, indicating that vortices are decaying at a rate \emph{faster} than would be expected for two-body collisions of vortices. However, given that our system has no boundaries, single-body loss is not possible, so vortices can only annihilate in pairs, meaning $\zeta<2$ should not be possible. We therefore conclude that the collisional model does not accurately describe the observed scaling in this system.

\subsection{Scaling of the momentum distribution \label{sub:momentum_spectrum}}

In renormalisation group theory, a many-body system far from equilibrium is predicted to show self-similar dynamical scaling when it passes by a nonthermal fixed point in the system space. Each fixed point is predicted to have associated scaling exponents, and the measurement of these exponents should allow for a universal classification of such fixed points across different physical systems~\cite{schmied_non-thermal_2019, mikheev_universal_2023}. For a Bose gas, this theoretical framework predicts a spatio-temporal scaling of the momentum distribution $n_k$ of the form
\begin{equation} \label{eq:nk_rescaling}
n_k(k, t) = (t/t^\prime)^\alpha n_k[(t/t^\prime)^\beta k, t^\prime],
\end{equation}
where $n_k(\textbf{k},t) \equiv |\hat{\psi}(\textbf{k},t)|^2$ and $\hat{\psi}$ is the Fourier transform of $\psi$. The reference time $t^\prime$ in this expression is an arbitrary time chosen within the scaling window, and $\alpha$ and $\beta$ are the two scaling exponents associated with the fixed point.

By taking the Fourier transform of Eq.~\eqref{eq:scaling_hypothesis} [assuming $G_\mathrm{eq}(r) \sim r^{-\eta}$, and noting that the Fourier transform of $G(\textbf{r},t)$ is identically $n_k(\textbf{k},t)$], for the 2D Bose gas we predict that these scaling exponents are related to the dynamical exponent $z$ via
\begin{equation}
\alpha = (d - \eta)/z, \hspace{2em} \beta=1/z,
\end{equation}
where $d=2$ is the dimensionality. Note that this expression includes a correction to the typically quoted relationship $\alpha/\beta=d$ for Bose gases~\cite{karl_strongly_2017, schmied_non-thermal_2019, mikheev_universal_2023}, resulting from the power-law decay of equilibrium correlations associated with BKT-type superfluidity in 2D.

\begin{figure}[t]
    \centering
    \includegraphics[width=\columnwidth]{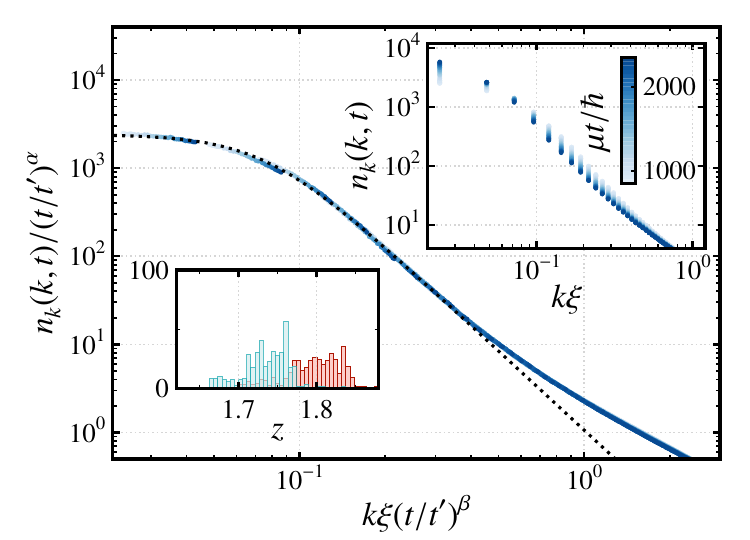}
    \caption{Evolution of the azimuthally-averaged particle number occupation spectrum $n_k(k,t)$ as a function of wavenumber, plotted for all times within the scaling window (normalised such that $\sum_\mathbf{k} n_k(\mathbf{k},t) = N$). The upper right inset shows the raw data, while the main frame shows the rescaled data, with exponents $\alpha=1.05$, $\beta=0.58$ used for the two axes. The lower left inset shows the distribution of $z_\alpha=(2 - \eta)/\alpha$ (red/right) and $z_\beta=1/\beta$ (cyan/left) values using different fitting regions. The black dotted line in the main frame shows the fitted function $f(\tilde{k}) = a / (1 + (\tilde{k}/\tilde{k}_0)^\kappa)$, with $\kappa=2.987$, $a=2440$, and $\tilde{k}_0=0.075$.}
    \label{fig:momentum_distribution}
\end{figure}

We measure $n_k(k,t)$ from our simulations, and fit our data to the universal form in Eq.~\eqref{eq:nk_rescaling}. The results are shown in Fig.~\ref{fig:momentum_distribution} over the temporal scaling window identified in Sec.~\ref{sub:correlation_length}. The raw data (upper right inset) collapse onto a single curve under rescaling (main frame), as predicted. Here, we have chosen the reference time to be the beginning of the scaling window, $t^\prime = 850 \, \hbar / \mu$. The data have been collapsed using the fitting procedure outlined in Ref.~\cite{karl_strongly_2017}, with the fitting region restricted to infrared wavenumbers satisfying $k \lesssim 0.75 \, \xi^{-1}$~\cite{karl_strongly_2017, chantesana_kinetic_2019}. We use the values of $\alpha$ and $\beta$ obtained from the fit to make two new measurements of the dynamical critical exponent, $z_\alpha = (2 - \eta) / \alpha$, and $z_\beta = 1/\beta$. As in Sec.~\ref{sub:correlation_length}, we use repeated fitting to obtain statistical distributions of these exponents, allowing us to estimate their uncertainty. The lower left inset of Fig.~\ref{fig:momentum_distribution} shows these two distributions, from which we measure $z_\alpha = 1.80(4)$ and $z_\beta = 1.74(3)$ [or equivalently, $\alpha=1.05(2)$ and $\beta=0.58(1)$]. These results are in reasonably good agreement with one another, as well as the measurements of $z$ from the previous two sections.

Typically, $\beta = 1/2$ is expected for defect-dominated scaling in a Bose gas, which is supported by numerical results in both 2D~\cite{karl_strongly_2017} and 3D~\cite{pineiro_orioli_universal_2015}. However, an `anomalous exponent' $\varepsilon$ is usually included in the definition of $\beta$, such that $\beta = 1/(2 - \varepsilon)$, allowing for deviations from this value (indeed, we have previously argued for an equivalent correction from the perspective of vortex dynamics~\cite{groszek_crossover_2021}). Whereas Ref.~\cite{karl_strongly_2017} obtained $\varepsilon$ consistent with zero for the 2D Bose gas (for scaling dominated by the `near-Gaussian' nonthermal fixed point), here we obtain $\varepsilon = 0.27(2)$. In Sec.~\ref{sec:Tdependence}, we explore the dependence of these exponents on the final temperature after the quench, finding that $\varepsilon$ reduces towards zero in the limit of low temperatures. Our quenches to lower temperatures are therefore consistent with the results of Ref.~\cite{karl_strongly_2017}.

As in earlier works (e.g.~Refs.~\cite{karl_strongly_2017, gazo_universal_2025}), we also fit the function $f(\tilde{k}) = a / (1 + (\tilde{k}/\tilde{k}_0)^{\kappa})$ to the rescaled momentum distribution in the region $\tilde{k} \equiv k\xi(t/t')^\beta \leq 0.3$, leaving $a$, $\tilde{k}_0$ and $\kappa$ as fitting parameters. The exponent $\kappa$ characterises the so-called Porod tail~\cite{bray_theory_1994, rutenberg_energy-scaling_1995}, which reflects the underlying structure of the disordered field on lengthscales $\xi \lesssim r \lesssim \Lc(t)$. In particular, $\kappa = d + 1 = 3$ has been predicted for sound wave turbulence~\cite{chantesana_kinetic_2019}, while $\kappa = d + 2 = 4$ is expected for randomly distributed vortices~\cite{bray_theory_1994, nowak_nonthermal_2012}. A value of $\kappa \approx 3$ was found in a recent experiment of coarsening in a 2D Bose gas~\cite{gazo_universal_2025}, which was attributed to the dominance of sound waves over vortices. Here we likewise obtain $\kappa = 2.987(6)$ (where the uncertainty denotes the fitting error; fit shown as a dotted black line in Fig.~\ref{fig:momentum_distribution}), suggesting the same interpretation. However, our simulations allow access to the vortex density (which was not measurable in Ref.~\cite{gazo_universal_2025}), which we have found to also exhibit dynamical scaling with a similar dynamical exponent $z$ to that measured from the overall field (see Sec.~\ref{sub:vortex_decay}). Our results therefore suggest that a Porod tail exponent of $\kappa \approx 3$ does not preclude the importance of vortices in the coarsening process, as it appears that both sound waves and vortices are contributing to the scaling in our simulations.

\subsection{Condensate growth and equilibrium correlations \label{sub:condensate_Geq}}

Early work on coarsening in a quenched conservative Bose gas applied finite-size scaling to extract the dynamical exponent $z$ and the equilibrium exponent $\eta$~\cite{damle_phase_1996}. This approach allows critical exponents to be calculated by comparing measurements of observables across different system sizes $L$, and extrapolating the behaviour to the thermodynamic limit $L \to \infty$. Using this method, Ref.~\cite{damle_phase_1996} obtained $z \sim 1$ in both 2D and 3D, which is significantly different from the results obtained here and in other recent works~\cite{karl_strongly_2017, groszek_crossover_2021}. To explore this apparent disagreement, we apply finite-size scaling analysis to our data.

As outlined in Ref.~\cite{damle_phase_1996}, the equilibrium correlation function should obey the (static) finite-size scaling relation
\begin{equation} \label{eq:G_scaling}
    G_{\rm eq}(r) = L^{-\eta} P \left( r/L \right),
\end{equation}
where $P$ is a universal scaling function satisfying $P(x \ll 1) = x^{-\eta}$ and $P(x \to 1) \to \mathrm{const}$. The equilibrium exponent $\eta$ quantifies the power-law decay of correlations in equilibrium, as described in Sec.~\ref{sub:correlation_length}. The condensate fraction, given by the fraction of atoms occupying the zero momentum mode $f_0(t) \equiv n_k(k=0,t)/N$, should also scale as
\begin{equation}  \label{eq:f0_scaling}
    f_0(t) = L^{-\eta} Q \left( t / L^z \right),
\end{equation}
where the scaling function $Q$ obeys $Q(x \gg 1) \to \mathrm{const}$. 

In order to make use of these finite-size scaling relations, we repeat our quench in four additional system sizes: $L \approx 16 \, \xi$ (with ensemble size $\mathcal{N}=1024$), $L \approx 33 \, \xi$ ($\mathcal{N}=512$), $L \approx 66 \, \xi$ ($\mathcal{N}=256$), and $L \approx 131 \, \xi$ ($\mathcal{N}=128$). We fix the energy- and particle- densities across all system sizes, along with the grid spacing and wavenumber cutoff. The initial states are generated in the same way as described in Sec.~\ref{sec:setup}, although we note that reducing $L$ leads to fewer momentum modes being present within the initially populated disk $|\textbf{k}|~\lesssim~0.92\,\xi^{-1}$.

\begin{figure}[t]
    \centering
    \includegraphics[width=\columnwidth]{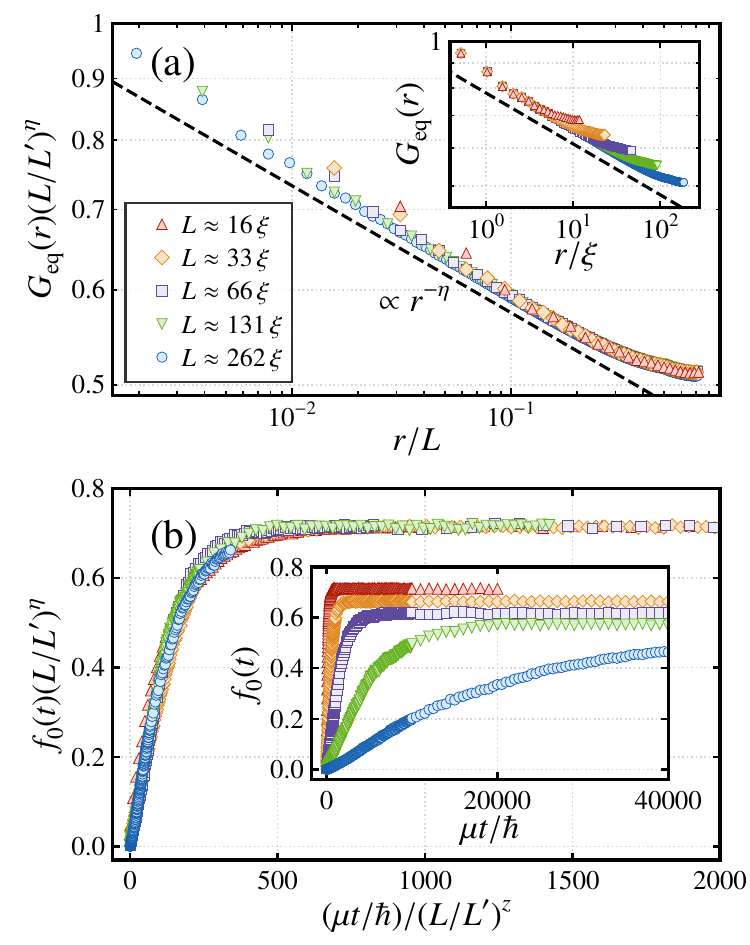}
    \caption{(a) Equilibrium correlation function $G_\mathrm{eq}(r)$ for five system sizes $L$, as identified in the legend. The data are shown both before (inset) and after (main frame) rescaling with system size $L$ according to Eq.~\eqref{eq:G_scaling}. A collapse is seen in the rescaled data for $r \gtrsim \xi$, and the power-law decay is well-described by the exponent $\eta=0.1058$ (dashed line) for $r \lesssim L/4$. (b) Evolution of the condensate ($k=0$ mode) fraction $f_0(t)$ for the same system sizes as in panel (a). The raw data are shown in the inset, while the main frame shows the collapse after rescaling according to Eq.~\eqref{eq:f0_scaling} with $z=1.8$. In (a) and (b), the reference system sizes are chosen to be $L' \approx 262 \, \xi$ and $L' \approx 16 \, \xi$, respectively.}
    \label{fig:f0Geq}
\end{figure}

The measured $G_\mathrm{eq}(r)$ and $f_0(t)$ for all system sizes are shown in the insets of Fig.~\ref{fig:f0Geq}(a) and (b), respectively. The condensate fraction is seen to plateau at late times to an equilibrium value, which we denote $\tilde{f}_0$. For $t \gg L^z$, the scaling form~\eqref{eq:f0_scaling} predicts that this value should vary with system size as $\tilde{f}_0(L) \sim L^{-\eta}$~\cite{nazarenko_bose-einstein_2014}. By fitting our $\tilde{f}_0(L)$ data with this power-law, we obtain $\eta = 0.1058(6)$ for this quench energy (data not shown).

With this measurement of $\eta$, it is straightforward to confirm the above scaling form, Eq.~\eqref{eq:G_scaling}. The measured $G_\mathrm{eq}(r)$ functions are shown in the main frame of Fig.~\ref{fig:f0Geq}(a) after rescaling both axes, and a convincing collapse is observed for $r \gtrsim \xi$. For ease of comparison with the raw data, we have chosen to rescale the vertical axis with respect to a reference system size $L'$. The power-law decay of correlations is evident, and excellent agreement is seen with the measured exponent $\eta$ (dashed line). For $r \gtrsim L/4$, finite-size effects begin to dominate, and algebraic decay is lost.

Using Eq.~\eqref{eq:f0_scaling}, we can also rescale $f_0(t)$, as shown in the main frame of Fig.~\ref{fig:f0Geq}(b).
By treating $z$ as a free parameter, the best collapse is obtained by eye for $z=1.8(1)$. However, we note that systematic variation is visible in the data before the late time plateau. This value of $z$ is significantly larger than that obtained in Ref.~\cite{damle_phase_1996} using this method, and is in agreement with the values obtained in the previous sections. We therefore conclude that the method can be used to extract the correct exponent, albeit with a significantly larger estimated uncertainty than the methods described in the previous sections.

\subsection{Decay of autocorrelation function \label{sub:autocorrelation}}

Finally, we explore the evolution of the two-time correlations after the quench. Universal scaling of these correlations has previously been demonstrated in related 2D systems such as XY and Ginzburg--Landau models (e.g.~\cite{liu_growth_1992, majumdar_growth_1995,lee_ordering_1995, nam_coarsening_2012}), as well as binary Bose gases~\cite{hofmann_coarsening_2014}, but to the best of our knowledge there have been no investigations into two-time correlations in the 2D scalar Bose gas.

We define the (magnitude of the) autcorrelation function
\begin{equation} \label{eq:autocorrelation_function}
A(t,t^\prime) = \left | \frac{\langle \psi^*(\textbf{r},t^\prime)\psi (\textbf{r},t) \rangle}{\sqrt{ \langle | \psi(\mathbf{r}, t^\prime) |^2 \rangle \langle | \psi(\mathbf{r}, t)| ^2 \rangle}} \right|,
\end{equation}
where the angular brackets here denote an average over both the spatial coordinate $\textbf{r}$ and the statistical ensemble. Taking the magnitude allows us to measure the temporal coherence between the field at times $t$ and $t'$, while ignoring any relative global phase accumulation~\cite{autocorrelation_footnote}. For $t \gg t^\prime$, it is predicted that this function should decay as a power-law in time,
\begin{equation} \label{eq:autocorrelation_scaling}
    A(t, t^\prime) \sim \left( \frac{\Lc(t)}{\Lc(t^\prime)} \right)^{-\lambda},
\end{equation}
i.e.~$A(t, t^\prime) \sim t^{-\lambda/z}$, with an exponent $\lambda$ that is independent of other dynamical exponents~\cite{fisher_nonequilibrium_1988, newman_new_1990, bray_theory_1994}.

\begin{figure}[b]
    \centering
    \includegraphics[width=\columnwidth]{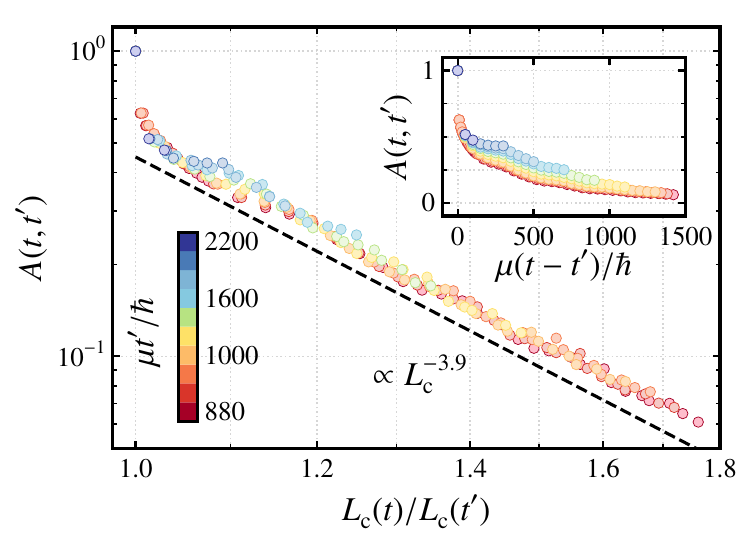}
    \caption{The evolution of the autocorrelation function $A(t, t^\prime)$ for multiple reference times $t^\prime$ during the scaling regime. The inset shows the decay as a function of time $t$ relative to the reference time $t'$. The main frame shows the same data plotted as function of the correlation length $\Lc(t)$, demonstrating a collapse with fitted power-law exponent $\lambda = 3.9(2)$. Note that reference times are sampled more finely for $\mu t'/\hbar < 1000$.}
    \label{fig:autocorrelation}
\end{figure}

The autocorrelation function measured from our simulations at $\epsilon \approx 2.8 \, \mu\xi^{-2}$ is presented in Fig.~\ref{fig:autocorrelation}. The inset displays $A(t,t')$ as a function of $t-t'$, for a range of reference times $t'$ within the scaling window (denoted by the colour scale). The correlations decay more slowly for later reference times, which can be attributed to the dynamics slowing down as the vortex density decreases. When the same data is plotted as a function of $\Lc(t)/\Lc(t')$, we find that it collapses onto a single universal curve (main frame of Fig.~\ref{fig:autocorrelation}). After a rapid initial decay, $A(t,t')$ begins to follow a power-law in $\Lc(t)/\Lc(t')$, with an exponent $\lambda = 3.9(2)$~\cite{lambda_footnote}, indicated by the dashed line. We obtain this exponent by fitting a power-law to each of our $t'$ datasets over the window $\Lc(t)/\Lc(t') \geq 1.1$ (where possible). The mean and standard deviation of the obtained $\lambda$ values then provide the exponent and its uncertainty, respectively.

In this analysis, we have restricted our measurement of the autocorrelation function to the regime where both $t$ and $t'$ fall within the scaling window determined in Sec.~\ref{sub:correlation_length}. Previous works have suggested that the exponent $\lambda$ may be different if $t'$ is chosen to be significantly before the scaling begins~\cite{sire_autocorrelation_2004, godreche_non-equilibrium_2004}. However, we find that choosing $t$ or $t'$ outside of the scaling window causes the scaling behaviour to break down.

\section{Temperature dependence of scaling exponents \label{sec:Tdependence}}

So far, we have explored the universal scaling behaviour of a quenched 2D Bose gas at a fixed energy density $\epsilon$, corresponding to a particular final temperature. However, theoretical arguments suggest that the post-quench temperature may play an important role in the coarsening of 2D systems, owing to the BKT form of superfluidity, which exhibits critical correlations for all temperatures below the BKT transition (i.e.~$G_\mathrm{eq} \sim r^{-\eta}$)~\cite{bray_theory_1994}. We therefore extend our analysis to a range of initial energies spanning the superfluid phase (corresponding to equilibrium exponents $0 \leq \eta \leq 0.25$) in order to characterise the coarsening as a function of microcanonical temperature.

To do this, we have repeated our simulations and analysis for a range of energy density values $\epsilon$ (omitting the finite-size scaling analysis of Sec.~\ref{sub:condensate_Geq}, as we found this to be the least precise method of measuring $z$). To reduce computation time, we have not evolved the PGPE to equilibrium at energy densities other than $\epsilon \approx 2.8 \, \mu/\xi^2$; instead, we use the stochastic projected Gross--Pitaevskii equation to obtain the equilibrium properties required for the analysis [specifically, $\eta$, $G_{\rm eq}(r)$ and $\tilde{n}_\mathrm{v}$; see Appendix~\ref{app:spgpe_equilibrium} for details]. For the two lowest $\epsilon$ quenches, we use a doubled system size ($L \approx 524 \, \xi$, with ensemble size $\mathcal{N}=64$), which we find to be necessary because $\Lc(t)$ reaches a larger value before scaling begins.

\begin{figure}[b!]
    \centering
    \includegraphics[width=\columnwidth]{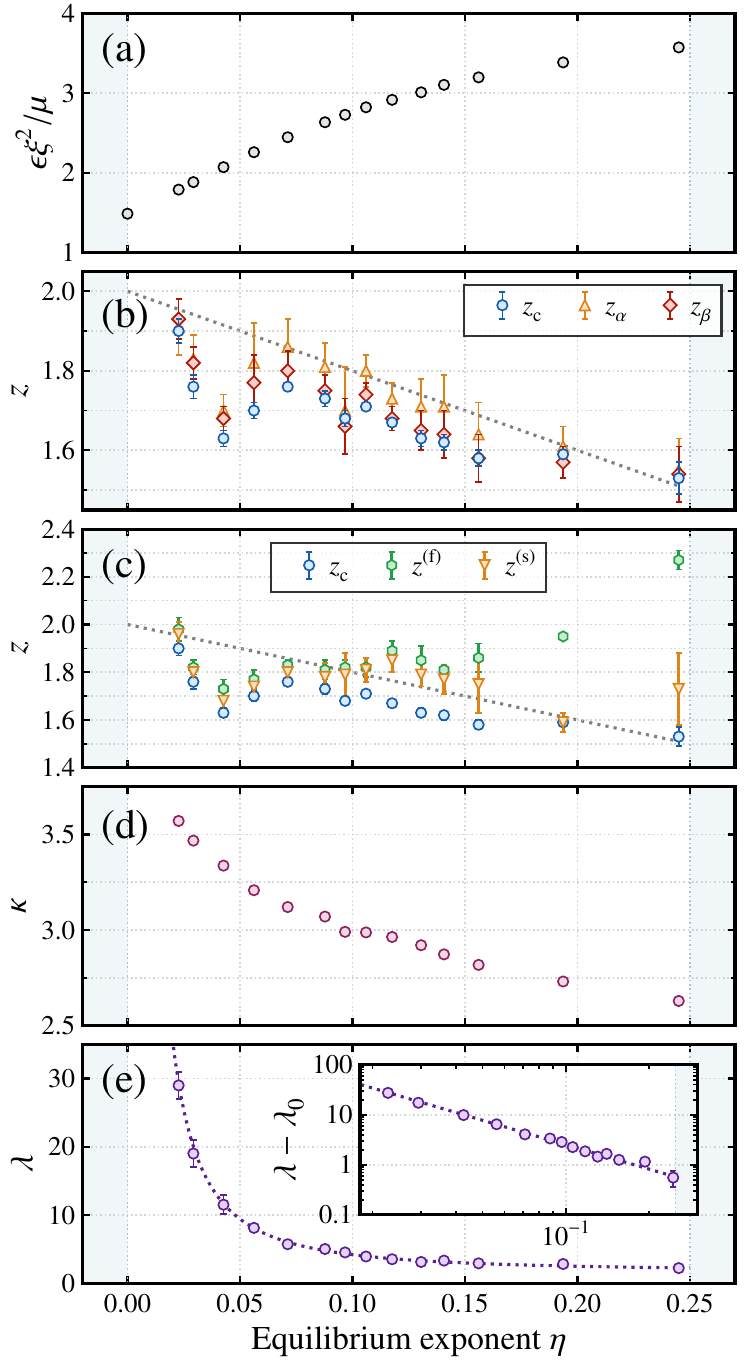}
    \caption{Universal scaling exponents as a function of the equilibrium exponent $\eta$, where $\eta=0$ and $\eta=0.25$ correspond to zero temperature and the BKT transition temperature, respectively (shading indicates inaccessible regions outside these values). (a) The correspondence between the chosen mean energy density $\epsilon$ and the measured equilibrium exponent $\eta$. (b) The dynamical exponent $z$, independently measured from the scaling of the correlation length ($z_\mathrm{c}$) and the momentum distribution [$z_\alpha = (d-\eta)/\alpha$ and $z_\beta = 1/\beta$]. (c) The dynamical exponent $z$ as measured from the filtered ($z^\mathrm{(f)}$) and subtracted ($z^\mathrm{(s)}$) vortex densities, with $z_\mathrm{c}$ included again for comparison. The dotted line in both (b) and (c) shows $2 - 2\eta$. (d) The Porod tail exponent $\kappa$. (e) The measured autocorrelation exponent $\lambda$, alongside a fit to Eq.~\eqref{eq:lambda_fit} (dotted line). The inset shows $\lambda-\lambda_0$ on a log--log axis to demonstrate the power-law scaling. The horizontal axis of the inset is the same as for the main frame.}
    \label{fig:z_vs_eta}
\end{figure}

The results of this analysis are presented in Figure~\ref{fig:z_vs_eta}. Figure~\ref{fig:z_vs_eta}(a) shows the relationship between the chosen energy density $\epsilon$ and the measured equilibrium exponent $\eta$ (the uncertainties in both variables are smaller than the point size). The quantitative relationship between these two variables is specific to our choice of parameters $gN$ and $\kcut$ in Eq.~\eqref{eq:PGPE}, as these determine the thermodynamic properties of the system at a given value of $\epsilon$. The exponent $\eta$, on the other hand, provides a model-independent measure of the superfluid density~\cite{prokofev_two-dimensional_2002}. Hence, if we treat $\eta$ as the control parameter (rather than $\epsilon$), we expect our results to be independent of the chosen interaction strength, particle number and wavenumber cutoff. We were unable to generate far-from-equilibrium states for $\eta \lesssim 0.02$ using our initialisation scheme, due to the highly constrained phase space at such low energies. For reference, we include the ground state ($\eta=0$) energy density, $\epsilon=gn^2/2 \approx 1.48 \, \mu\xi^{-2}$, in the plot.

Figure~\ref{fig:z_vs_eta}(b) shows the dynamical exponent $z$ measured from three observables: the growth of the correlation length, $\Lc \sim t^{1/z_\mathrm{c}}$ (as in Sec.~\ref{sub:correlation_length}), as well as the two exponents from the momentum distribution scaling, $z_\alpha$ and $z_\beta$ (as in Sec.~\ref{sub:momentum_spectrum}). The three measures are in good agreement with one another for all $\eta$, although there does seem to be a weak systematic bias, $z_\alpha \gtrsim z_\beta \gtrsim z_\mathrm{c}$, albeit within overlapping error bars. This may be attributable to, e.g., the choice of scaling function threshold $F_0$ (for $z_\mathrm{c}$) and the choice of wavenumber fitting window for $n_k(k,t)$ (for $z_\alpha$ and $z_\beta$). We also note the surprising nonmonotonic behaviour of $z$ at low $\eta$, which appears consistently across all measurements of $z$. It is not clear what could cause such a feature. It may be due to systematic effects---such as the increasingly constrained initial conditions in the low temperature limit---although this seems unlikely, as there is no indication that the $\kappa$ and $\lambda$ exponents are also affected [see Figs.~\ref{fig:z_vs_eta}(d,e), and discussion below]. Additionally, we find the scaling function $F$ in Eq.~\eqref{eq:scaling_hypothesis} [and Fig.~\ref{fig:correlation_function}(a)] to be independent of temperature, indicating that the spatial correlations in the field approach equilibrium in a universal way for all $\eta$, despite the varying value of $z$. This leaves open the possibility that the nonmonotonicity in $z$ is a genuine physical effect, potentially resulting from a competition between vortices and waves in the coarsening as $\eta$ is varied. Nonetheless, we find that $z<2$ in general, with an overall temperature dependence: $z \approx 1.9$ for our lowest energy quench, decreasing (on average) to $z \approx 1.5$ near the BKT transition. Similar behaviour is observed in the dynamical exponents measured from the filtered ($z^\mathrm{(f)}$) and subtracted ($z^\mathrm{(s)}$) vortex densities at low $\eta$ [Fig.~\ref{fig:z_vs_eta}(c)], although these measurements become unreliable above $\eta \gtrsim 0.1$ due to the rapid growth of the thermally activated dipole density~\footnote{For the filtering, we monotonically vary the radius between $k_\mathrm{f} \approx 0.92 \, \xi^{-1}$ (for the lowest temperature quenches) to $k_\mathrm{f} \approx 0.46 \, \xi^{-1}$ (for the highest temperatures).}.

In Ref.~\cite{groszek_crossover_2021}, we argued that $z = 2 - \varepsilon$ for the conservative 2D Bose gas, where the correction $\varepsilon$ arises from a vortex mobility $\mu_\mathrm{v}$ that varies as a power-law with the correlation length, $\mu_\mathrm{v} \sim \Lc^{\varepsilon}$. For $\varepsilon = 0$, the mobility is constant, and point-vortex dynamics with $z=2$ are recovered~\cite{groszek_crossover_2021}. The results in Fig.~\ref{fig:z_vs_eta}(b,c) suggest that this limit is almost realised at the lowest temperatures, but that $\varepsilon$ would, in general, be temperature dependent. In fact, the general trend in our results roughly follows $\varepsilon \approx 2\eta$ [dotted line in panels (b,c)], suggesting a possible relationship between these two exponents. Physically, vortices move in gradients of both the phase and density of $\psi$~\cite{tornkvist_vortex_1997, groszek_motion_2018}, so it seems reasonable that the vortex mobility may change as the collective excitations of the field grow in amplitude at higher temperatures. 

The Porod law exponent $\kappa$ [Fig.~\ref{fig:z_vs_eta}(d)] provides further insight into the interplay between vortices and waves. It is found to smoothly vary from $\approx 2.6$ at the highest temperature sampled to $\approx 3.6$ at the lowest. The trend $\kappa \to \sim 4$ suggests that vortices are becoming an increasingly dominant feature of the coarsening process at low temperatures. Surprisingly though, the high temperature values of $\kappa$ fall even below the prediction for weak-wave turbulence, $\kappa=3$ (while remaining significantly above the equilibrium value $\kappa = 2 - \eta$, corresponding to equipartition of energy). We are not aware of any predictions for $\kappa < 3$ in this system, although we note that this anomalously low value---obtained for $\eta \gtrsim 0.1$---coincides with the proliferation of thermal dipoles (and the resulting departure of $z^\mathrm{(f)}$ and $z^\mathrm{(s)}$ from $z_\mathrm{c}$). It may be that these dipoles can in fact modify the structure of the field at intermediate lengthscales once their density is sufficiently high, causing this reduction in $\kappa$. Further investigation into the physical mechanisms giving rise to $\kappa < 3$ is left as an avenue for future work.

Finally, we plot the autocorrelation exponent $\lambda$ as a function of $\eta$ in Fig.~\ref{fig:z_vs_eta}(e). For the highest temperature sampled ($\eta \approx 0.245$), we find $\lambda =2.2(2)$, consistent with previous general predictions that $\lambda = d = 2$ should hold at the critical point~\cite{majumdar_growth_1995}, and close to a previous numerical measurement $\lambda=2.21$ obtained from the conservative XY model~\cite{nam_coarsening_2012}. However, as the temperature is reduced, we observe a rapid increase in $\lambda$, which appears to diverge at the lowest temperatures sampled~\footnote{At low temperatures, the autocorrelation function decays so quickly that it is equally well described by exponential (rather than power-law) decay. However, it is unclear why the scaling would change from power-law at high temperatures to exponential at low temperatures.}. Surprisingly, we find that the exponent is well described by the relation:
\begin{equation} \label{eq:lambda_fit}
    \lambda = c \eta^{-\sigma} + \lambda_0.
\end{equation}
Fitting this expression to the data gives an exponent $\sigma=1.62(8)$, an offset $\lambda_0=1.6(2)$, and a prefactor $c=0.06(2)$. The fit is shown as a dotted line in the main frame of Fig.~\ref{fig:z_vs_eta}(e). To emphasise how well the data is described by the power-law component of this fit, we plot $\lambda - \lambda_0$ as a function of $\eta$ in the inset of Fig.~\ref{fig:z_vs_eta}(e). The resulting data shows convincing power-law scaling, with $\lambda - \lambda_0$ varying over $\sim 2$ orders of magnitude.

We have found no other literature that reports a scaling exponent that is itself described by a power-law function. Intuitively, it would seem reasonable to attribute the decay of the autocorrelation function to thermal fluctuations, which reduce phase coherence in the field, and should serve to wash out the system's memory. However, our results clearly show that this is not the case, since $\lambda$ is largest in the low-temperature limit. Instead, it appears that the memory of the system is determined primarily by the vortex dynamics. In particular, we observe a much slower vortex annihilation rate in the lowest temperature simulations, which allows the field to significantly reconfigure itself (via vortex--vortex interactions) between annihilation events. This, in turn, leads to a much greater decrease in $A(t,t')$ for a given increase in $\Lc$, and therefore a larger exponent $\lambda$.

\section{Conclusions \label{sec:conclusion}}

We have explored the universal coarsening dynamics of a two-dimensional Bose gas undergoing conservative evolution following a quench. We find that the dynamical exponent $z$ depends on the initial energy of the system, and on average decreases for quenches to higher energies. This behaviour is consistent across a range of independently measured observables. The value $z \approx 2$ typically expected for this system only appears to be valid in the low temperature limit, while $z \approx 1.5$ is obtained at the highest temperatures for which the system still exhibits superfluidity. The Porod tail exponent $\kappa$ also exhibits temperature dependence, varying monotonically with temperature in the range $2.6 \lesssim \kappa \lesssim 3.6$, which points to varying contributions from vortices and sound waves to the coarsening process. Likewise, we find the autocorrelation exponent $\lambda$ to be strongly temperature dependent: while $\lambda \approx 2$ for quenches to just below the BKT transition (in agreement with earlier predictions), $\lambda$ appears to diverge at low temperatures. Surprisingly, this exponent itself is well described by a power-law $\lambda \sim \eta^{-\sigma}$, with the new exponent measured to be $\sigma=1.62(8)$.

Taken together with the results of our previous work~\cite{groszek_crossover_2021}, we have found that the nonequilibrium scaling exponents in the 2D Bose gas vary significantly with global system properties such as the initial energy and the strength of any added dissipation. In fact, the vortex configuration itself also appears to play an important role in the coarsening; same-sign clustering of vortices has been found to be associated with an anomalously high $z \approx 5$ and $\kappa \approx 6$~\cite{karl_strongly_2017}, while tightly bound dipoles give $2 \lesssim z \lesssim 4$ (with dissipation included)~\cite{tattersall_out_2025, chu_quenched_2001, forrester_exact_2013}. These results complicate the typical simple picture that $z=2$ in this system (potentially with logarithmic corrections~\cite{yurke_coarsening_1993, rutenberg_energy-scaling_1995}), and suggest that the dynamics may not adhere to a simple characterisation in terms of a single nonequilibrium universality class. This may be attributable to the BKT physics particular to 2D superfluids---which introduces corrections to scaling via the equilibrium exponent $\eta$ and the presence of thermal vortex dipoles---or it may be a more general feature of coarsening at high temperatures, which introduces a competition between defects and collective excitations.

\begin{acknowledgments}
We thank Paolo Comaron, Martin Gazo, Nikolaos Proukakis, Tapio Simula and Lewis Williamson for useful discussions. We acknowledge financial support from the UK EPSRC [grant number EP/R021074/1] (AJG and TPB) and the Australian Recearch Council Centre of Excellence for Engineered Quantum Systems [project number CE170100009] (AJG). This research made use of the Rocket High Performance Computing service at Newcastle University.

Data supporting this publication are openly available under a Creative Commons CC-BY-4.0 License found in Ref.~\cite{data}.
\end{acknowledgments}


\appendix

\section{Determining equilibrium properties \label{app:spgpe_equilibrium}}

Due to the long evolution time required to reach equilibrium in the PGPE, we opt to use the stochastic projected Gross--Pitaevskii equation (SPGPE),
\begin{equation} \label{eq:SPGPE}
    \diff \psi = \mathcal{P} \biggl \lbrace - \frac{i}{\hbar} L_{\rm GP} \psi \diff t + \frac{\gamma}{\hbar} ( \mu - L_{\rm GP}) \psi \diff t + \diff W \biggr \rbrace,
\end{equation}
to sample the equilibrium behaviour for all data points in Fig.~\ref{fig:z_vs_eta} except $\epsilon \approx 2.8\,\mu\xi^{-2}$. Here, the operator $L_{\rm GP} = - (\hbar^2 / 2m) \nabla^2 + g |\psi| ^2$. In this model, the field $\psi$ is in connected to a thermal reservoir at temperature $T$ and with chemical potential $\mu$. The dimensionless dissipation rate $\gamma$ determines how strongly the system is coupled to the reservoir. The choice of $\gamma$ does not affect the equilibrium behaviour~\cite{gardiner_stochastic_2003}, so we set $\gamma=1$ without loss of generality. The complex Gaussian noise $\diff W(\textbf{r},t)$ has variance $\langle \diff W^* (\textbf{r},t) \diff W (\textbf{r}',t) \rangle = (2 \gamma k_{\rm B} T / \hbar) \delta (\textbf{r} - \textbf{r}') \diff t$.

The SPGPE samples a grand canonical ensemble, and hence to match it most closely to the microcanonical PGPE for a particular mean energy density $\epsilon$ and particle density $n$, we must choose $\mu$ and $T$ such that $\langle n \rangle_\mathrm{SPGPE} \approx n_\mathrm{PGPE}$ and $\langle \epsilon \rangle_\mathrm{SPGPE} \approx \epsilon_\mathrm{PGPE}$, with the averages taken over statistically equivalent samples. Numerically, we are able to achieve these equivalences to within $\lesssim 1 \%$ precision for all chosen energy densities.

Using a spatially uniform initial condition $\psi = (\mu / g)^{1/2}$, we find that the system equilibrates by $\approx 200 \, \hbar / \mu$ under evolution of Eq.~\eqref{eq:SPGPE}. For each energy density $\epsilon$ sampled in Fig.~\ref{fig:z_vs_eta}, we run an ensemble of 64 SPGPE simulations, and uniformly sample ten equilibrium microstates from each within the temporal window $200 \lesssim \mu t / \hbar \lesssim 300$. We then average this data to obtain the equilibrium correlation function $G_{\rm eq}(r)$ and corresponding algebraic decay exponent $\eta$, as well as the mean density of thermal vortices $\tilde{n}_\mathrm{v}$ (see Sec.~\ref{sub:vortex_decay}).

To compare the observables obtained from the PGPE and SPGPE, we have sampled the equilibrium behaviour for the energy density $\epsilon \approx 2.8 \, \mu \xi^{-2}$ considered in Sec.~\ref{sec:quench_example} (at system size $L= 262 \, \xi$) using both methods. Power-law fits to the correlation function $G_{\rm eq}(r)$ over the range $10 \leq r/\xi \leq 64$ give decay exponents $\eta_{\rm PGPE} \approx 0.108$, $\eta_{\rm SPGPE} \approx 0.105$, demonstrating that the two approaches are in close agreement. Likewise, the equilibrium vortex densities $\tilde{n}_\mathrm{v,PGPE} = 2.3(3) \times 10^{-3} \, \xi^{-2}$ and 
$\tilde{n}_\mathrm{v,SPGPE} = 2.0(2) \times 10^{-3} \, \xi^{-2}$ overlap within uncertainty.


\begin{thebibliography}{88}
\bibitem [{aut()}]{autocorrelation_footnote}%
  \BibitemOpen
  \href@noop {} {}\bibinfo {note} {We find that the autocovariance function
  [obtained by replacing $\psi(\mathbf{r},t) \to \psi(\mathbf{r},t) - \langle
  \psi(\mathbf{r},t)\rangle_\mathbf{r}$ in
  Eq.~\eqref{eq:autocorrelation_function}] behaves almost identically to the
  autocorrelation function, since $|\langle \psi(\mathbf{r},t)
  \rangle_\mathbf{r}| \ll 1$ in the scaling regime.}\BibitemShut {Stop}%
\end{thebibliography}

%

\end{document}